\documentstyle[12pt,psfig]{article}
\newcommand{\drawsquare}[2]{\hbox{%
\rule{#2pt}{#1pt}\hskip-#2pt
\rule{#1pt}{#2pt}\hskip-#1pt
\rule[#1pt]{#1pt}{#2pt}}\rule[#1pt]{#2pt}{#2pt}\hskip-#2pt
\rule{#2pt}{#1pt}}
\newcommand{\Yfund}{\raisebox{-.5pt}{\drawsquare{6.5}{0.4}}}

\def\be{\begin{equation}}
\def\ee{\end{equation}}
\def\bc{\begin{center}}
\def\ec{\end{center}}
\def\bea{\begin{eqnarray}}
\def\eea{\end{eqnarray}}

\def\nn{\nonumber}
\def\la{\langle}
\def\ra{\rangle}

\def\det{{\rm det}}

\def\ov{\overline}
\def\qbar{\ov{q}}
\def\QQbar{\ov{Q}}

\def\tr{{\rm Tr}}
\begin{document}
\setcounter{footnote}{1}

\begin{titlepage}

\begin{flushright}
UCB-PTH-01/42\\
hep-th/0112082\\
\end{flushright}

\vskip 2cm
\begin{center}
{\large \bf New Theories With Quantum Modified Moduli Space}

\vskip 1.2cm 

Hitoshi Murayama$^{1,2}$ and Elena~Perazzi$^{2}$
\vskip 0.4cm

{\it $^1$ Department of Physics\\
University of California, Berkeley, California 94720}\\
{\it $^2$ Theoretical Physics Group\\
Ernest Orlando Lawrence Berkeley National Laboratory, MS 50A-5101\\
University of California, Berkeley, California 94720}\\
\vskip 0.2cm
\vskip 2cm
(\today)
\abstract{
Kutasov--type duals of supersymmetric gauge theories had been studied
only in the dual regime and the s-confining case.  Here we extend the
discussion to the case of less flavor, analogous to the case of
quantum-modified moduli space in Seiberg duality.  Unlike the Seiberg
duality, however, we find that parts of the moduli space become
superconformal, generalizing the so far isolated example of $SU(2)$
theory with two doublets and a triplet.  We also point out that the
magnetic superpotential needs to be augmented by an additional
instanton-generated piece when the magnetic group is $SU(2)$.}
\end{center}
\end{titlepage}

\setcounter{footnote}{0}

\section{Introduction}
%

Many of recent advances on understanding gauge theories and string
theories rely on the notion of duality.  The concept of duality varies
depending on the context.  In the case of string dualities, they are
supposed to be exact equivalence between two theories.  Two theories
are different descriptions of the same theory valid in different
limits.  In the case of field theory dualities, they mostly refer to
the equivalence of low-energy limits, a generalization of the concept
of universality class in critical phenomena.  Supersymmetric gauge
theories have proven to be wonderful testing grounds for various types
of dualities.

Among dualities in supersymmetric gauge theories, Kutasov--type
dualities \cite{kut1,kut2} offer the most complex and rich phenomena
of duality.  The original version by Kutasov--Schwimmer is a duality
between $SU(N)$ gauge theory with one adjoint chiral multiplet $X$ and
$F$ quark favors, supplemented by a superpotential of the form $W = h
\lambda \tr X^{k+1}$ and an $SU(kF-N)$ theory with a similar particle
content but more complicated superpotential.  Furthermore, a later
paper together with Seiberg \cite{kut3} demonstrated a beautiful
structure of chiral rings, discussed originally in the context of
two-dimensional superconformal theories.  This type of duality had
later been extended to many other gauge groups and both symmetric and
anti-symmetric tensor fields \cite{more}.  They also offer the most
non-trivial examples of discrete anomaly matching \cite{discrete}.

Despite the successes mentioned above, Kutasov--type dualities still
suffer from the lack of a complete picture.  For instance, the dynamics
in the absence of the superpotential is still a complete mystery.  For
smaller number of flavors, we expect analogues of $F \leq N+1$
cases of Seiberg's $SU(N)$ dualities; however the only case that had
been worked out in full detail is the case of $kF-N=1$, an
analogue of $F = N+1$ leading to new s-confining theories.
Indeed, a free theory of composites with an irrelevant superpotential
was found.  

The purpose of this paper is to investigate the dynamics of the
$SU(N)$ Kutasov--type duals when $F = k N$.  Naively, this gives an
analogue of $F=N$ case in Seiberg duality with a quantum-modified
moduli space.  We show that such a picture is almost correct, but with
an important difference.  At certain sections of the quantum-modified
moduli space, there appear interacting non-trivial superconformal
theories.  The closest example that exhibits a similar behavior
studied in the literature so far is the $SU(2)$ theory with two
doublets and a triplet \cite{isphases}.  Therefore our work
generalizes this isolated example to a whole class of theories.

The paper is organized as follows.  In Section 2, we review the known
results from Kutasov--Schwimmer and later with Seiberg on $SU(N)$
gauge theory with $F$ flavors and an adjoint $X$.  This result applies
only when $kF-N > 1$.  In Section 3, we study the electric theory with
$kF=N$. In Section 4 we study the corresponding magnetic theory 
by
decoupling a flavor from the $SU(k)$ theory with $F+1$ flavors.
We draw our conclusions 
in Section 5.
Additional checks of Kutasov duality are considered in the Appendix.
There we also discuss a subtlety of the case $k=2$, $2F-N=2$:
it turns out that the usual Kutasov superpotential for the magnetic
theory is incomplete in that case, and an additional term must be added
to reproduce the result of the electric theory.

\section{Overview of the theory and its dual}

\subsection{Known Results}

In this section, we review the known Kutasov--type duals.
Consider a $SU(N) $ gauge theory with 
$F$ flavors in fundamental ($Q$) and anti-fundamental ($\QQbar$) 
representation and an adjoint ($X$), and with a superpotential
\be
\label{superpot}
W=\frac{h}{k+1} \ \tr X^{k+1},
\ee
where $h$ is a coupling constant of dimension $k-2$. 
The symmetry properties of the matter fields are taken to be:

\begin{eqnarray}
\label{spectrumel}
\begin{array}{c|c|cccc}
& SU(N) & SU(F) & SU(F) & U(1) & U(1)_R \\ \hline
X & Adj & 1 & 1 & 0 & \frac{2}{k+1} \\
&&&&&\\
Q & \Yfund & \Yfund & 1 & 1 &  1-{2N\over (k+1)F} \\
&&&&&\\
\ov{Q} & \overline{\Yfund} & 1 & \ov{\Yfund} & -1 &   1-{2N \over (k+1)F}
\end{array} 
\end{eqnarray}

If we impose the $D$-flatness conditions and mod out by gauge
transformations (or, equivalently, mod out by complexified gauge
transformations), we can transform $X$ to a Jordan normal form
\be
\label{jordan}
X=\left(\begin{array}{ccccccccc}
       a & 1 &    &     &     &     &  &  &  \\
         & a & 1  &     &     &     &  &  &  \\
         &   & a  &     &     &     &  &  & \\
         &   &    &  b  &  1  &     &  &  &        \\
         &   &    &     &  b  &     &  &  &    \\
         &   &    &     &     &  c  &  &  &  \\
         &   &    &     &     &     &. &  &  \\
         &   &    &     &     &     &  & . &  \\
         &   &    &     &     &     &  &  & z \\
\end{array} \right)
.
\ee
The moduli space of the theory is then obtained 
by imposing the $F$-flatness condition from (\ref{superpot}):
\be
\label{f-flat}
X^k-{1 \over N} {\rm Tr} X^k=0.
\ee
This way, we end up with two possibilities: either $X$ is diagonal and
it is $X^k=v^k I$, where $v$ is an arbitrary complex number and $I$ is
the identity matrix, or it has all zero diagonal entries (and thus it
is a singular matrix with vanishing $k$th power and the $D$-term is
canceled by corresponding contributions from quarks).  We can think
of the latter case as the origin of the flat direction $v$ in moduli
space.

For $v\neq 0$, vacua of the gauge theory can be labeled by 
sequences of integers $(r_1,r_2,...,r_k)$, with $\sum r_i=N$,
where $r_i$ is the number of eigenvalues of $X$ residing 
in the $i-$th of the $k$ roots of $v^k$.
The gauge group is broken by the $X$ expectation value to
\be
\label{elfact}
SU(N) \rightarrow SU(r_1)\times SU(r_2)\times ... \times SU(r_k)\times
U(1)^{k-1}.  
\ee 
Obviously, $\sum_{i=1}^k r_i = N$.  
Thus, at low energies we are left with $k$ decoupled QCD theories with
gauged baryon number(s).  The quantum behavior of each $SU(r_i)$ is known
from the pioneering work by Seiberg \cite{dual}.  In particular, remember
that classical vacua with $r_i\leq F$ are removed from the quantum
moduli space (so, at least for $v\neq 0$, theories with the matter
content of Table~1 can have stable vacua at all only if $kF\geq N$).
For $r_i> F$ the quantum moduli space is identical to the classical
one, and for $r_i=F$ the two spaces are different, the classical
(compositeness) relation between mesons and baryons, $\det M^{(i)} -
B^{(i)}\ov{B}^{(i)}=0$, being replaced by the quantum one: 
\be
\label{qmc1}
\det M^{(i)} - B^{(i)}\ov{B}^{(i)}=\Lambda^{(i)3r_i-F}.
\ee

Moreover, remember that there are strong pieces of evidence
\cite{dual} that a $SU(r)$ theory with $F$ flavors is dual to a
$SU(F-r)$ theory with $F$ flavors, one singlet and an appropriate
tree-level superpotential.  As a natural generalization of this
duality conjecture, Refs.~\cite{kut1,kut2,kut3} suggested that the
theory with the field content of Table~1 is dual to a theory with
$SU(kF-r)$ gauge group, $F$ flavors of (dual) quarks ($q$) and
antiquarks ($\qbar$), $k$ singlets $M_j$ and an adjoint $Y$, with the
symmetry properties

\begin{equation}
\begin{array}{c|c|cccc}
& SU(kF-N) & SU(F) & SU(F) & U(1) & U(1)_R \\ \hline
Y & Adj & 1 & 1 & 0 & {2\over (k+1)}\\
&&&&&\\
q & \Yfund & \overline{\Yfund} & 1 & {N \over kF-N} &  1-{2(kF-N)
  \over (k+1)F} \\  
&&&&&\\
\ov{q} & \overline{\Yfund} & 1 & \Yfund & - {N \over kF-N}
&1-\frac{2(kF-N)}{(k+1)F} \\
&&&&&\\
M_j & 1 & \Yfund & \ov{\Yfund} & 0 & 2-{4N\over (k+1)F}+{2\over (k+1)} (j-1) 
\end{array}
\end{equation}

Except for the case $k=2$, $2F-N=2$, 
the magnetic tree-level superpotential is taken to be
\be
\label{magn}
W_{magn}=-\frac{h}{k+1} {\rm Tr}Y^{k+1} 
+ {h \over \mu^2} \sum_i M_i \qbar Y^{k-i} q.
\ee
The $F$-flatness condition for $Y$ is similar in form to that of
eq.(\ref{f-flat}), therefore the magnetic moduli space also contains
a flat direction $v$, analogous to that of the electric theory.
Points in the electrical moduli space where the $SU(N)$ theory splits
into the product of the $k$ $SU(r_i)$ theories correspond to points in
the magnetic moduli space where the magnetic $SU(kF-N)$ theory splits
into the product of the corresponding dual $SU(F-r_i)$ theories.
Notice that points in the classical electric moduli space
with some $r_i<F$ are removed from the corresponding
magnetic moduli space because $SU(F-r_i)$ cannot exist then. Thus, the
two spaces do not agree classically, but only quantum mechanically.

In the case $k=2$, $2F-N=2$, Tr$Y^3=0$, and, as shown in the Appendix,
the magnetic superpotential requires an additional term, and is thus
\be
\label{additional}  
W_{magn}={h \over \mu^2} \sum_i M_i \qbar Y^{k-i} q
-\left({{\rm det} M^{(1)}\over  \Lambda^{(1)\ 3N/2-(F+1)}}+
{{\rm det} M^{(2)}\over  \Lambda^{(2)\ 3N/2-(F+1)}}\right),
\ee
where 
\be
M^{(1),(2)}= {M_1\pm v  M_2 \over 2v},
\ee
and, consistently with the general case, $v=\sqrt{{1 \over 2 }{\rm Tr} Y^2}$.
  
At the origin $v=0$, the adjoint doesn't
decouple from the low-energy theory.
and the moduli space can be described at all energies by generalized mesons
\be
\label{genmes}
(M_j)^i_{\ov{i}}=\QQbar_{\ov{i}} X^{j-1}Q^i;\ \ \ j=1,...,k;\ \
i,\ov{i}=1,...,F\ , 
\ee
baryons
\be
\label{genbar}
B^{(n_1,n_2,...,n_k)}=Q^{n_1}(XQ)^{n_2}....(X^{k-1}Q)^{n_k}
\ee
and, finally, Tr$X^j$ with $j=1,...,k$.
The mesons (\ref{genmes}) can be also thought of as blocks of the matrix
\be
\label{mes1}
\left(\begin{array}{ccccc}
       \QQbar Q & \QQbar XQ  &...  & \QQbar X^{k-2} Q  & \QQbar X^{k-1} Q \\
       \QQbar X Q  & \QQbar X^2 Q  &...  & \QQbar X^{k-1} Q      & 0  \\
        .   &     &   &    &        \\
        .   &     &   &    &        \\
        .   &     &   &    &         \\
       \QQbar X^{k-1}Q   & 0   & ... & 0 & 0    
\end{array} \right)
\ee
constructed from the ``dressed'' quarks and anti-quarks
\be
\label{dressed}
Q_{(l)}=X^{(l-1)}Q;\ \ \QQbar_{(l)}=X^{(l-1)}\QQbar;\ \ l=1,...k\ \ .
\ee 

A mapping between the above gauge-invariant operators and those of the
magnetic dual can be established as follows: the mesons in
eq.(\ref{genmes}) correspond to the elementary singlets of the
magnetic theory. The correspondence between the electric baryons ($B$)
defined in eq.(\ref{genbar}) and magnetic ones ($b$), constructed in
an analogous way out of $q$ and $Y$, is
\be
B^{(n_1,n_2,....,n_k)} \leftrightarrow  b^{(m_1,m_2,....,m_k)},\ \ \
m_l=F-n_{k+1-l},\ \ l=1,...,k 
\ee
and the traces Tr$X^j$ are simply mapped to the analogous $-$Tr$Y^j$.
 
Non-trivial tests of the duality conjecture include:
\begin{itemize}
\item The fact that the charge assignment of Table~1 and Table~2,
  necessary for 't Hooft anomaly matching, is also compatible with the
  above mapping.
\item The fact that the moduli space of the electric and magnetic
  theory are equivalent after including instanton effects as we show
  in the Appendix. 
\item The fact that the duality is preserved under mass deformations.
\end{itemize}
The discussion in Refs.~\cite{kut1,kut2,kut3} established a picture of
the IR behavior of the theory at $v=0$ for all the values of $F$ such
that $kF-N>1$: the theory is in the free electric phase for $F>2N$, in
the free magnetic case for $F<{2\over 2k-1} N$ and in the non-Abelian
Coulomb phase for the values of $F$ in the intermediate range.

Furthermore,  Csaki and one of the authors (HM) studied the case
$kF-N=1$ \cite{cm}, by adding a mass deformation to the case
$kF-N=k+1$. They found that for $kF-N=1$ the theory is always
confining, and obtained explicitly the confining superpotential as a
$k$-instanton effect in the magnetic theory.

Along the same line, we want to investigate whether duality can
elucidate the behavior of the theory for $kF=N$.  In section 3 we will
analyze in detail the electric theory and obtain the quantum
modification of the moduli space in the limit $v\rightarrow 0$.  This
quantum modification will be obtained explicitly as an instanton
effect in section 4, in the framework of the dual magnetic theory.

\section{The Electric theory with {\bf $kF=N$}}

Here we investigate the $SU(kF)$ theory with $F$ pairs of quarks $Q$, 
$\ov{Q}$ and an adjoint field $X$ with the superpotential
\begin{equation}
    W = {h\over k+1} {\rm Tr} X^{k+1} .
\end{equation}
The classical moduli space is given in terms of the mesons $M_{j} = \ov{Q} 
X^{j-1} Q$ $(j = 1, 2, \cdots, k)$, baryons $B = Q^{F} 
(XQ)^{F} \cdots (X^{k-1} Q)^{F}$, $\ov{B} = \ov{Q}^{F} 
(X\ov{Q})^{F} \cdots (X^{k-1} \ov{Q})^{F}$, ${\rm Tr}X^{j}$, $j=2,3,...,k$.
If we define the matrix ${\cal M}$ such that ${\cal M}_{ij}=M_{i+j-1}$
for $i+j\le k+1$ and ${\cal M}_{ij}=v^k M_{i+j-(k+1)}$ otherwise,
baryons and mesons are subject to the constraint
\begin{equation}
\label{constraint}
    \det {\cal M} - B\ov{B}=0.
\end{equation}
In the limit $v\rightarrow 0$ {\cal M} reduces to the matrix in 
eq.(\ref{mes1}) and the constraint (\ref{constraint}) 
simplifies to
\be
\label{constr1}
(-1)^{{k(k-1)\over 2}} {\rm det} (M_k)^k -B\ov{B}=0.
\ee

Let us briefly discuss the number of degrees of freedom of the
classical moduli space.  We have $k$ $M_{j}$'s each with $F^{2}$
components, $B$ and $\ov{B}$, subject to the constraint
(\ref{constraint}), plus the traces ${\rm Tr}X^{j}$, $j=2,3,...,k$.
Therefore, we have $kF^{2}+k$ dimensions.

Along the $F$-flat and $D$-flat direction $X^{k} = v^{k} I$ with $v
\neq 0$, $X$ takes the form $X = v\, {\rm diag}(\underbrace{1, \cdots,
  1}_{n_{1}}, \underbrace{\omega, \cdots, \omega}_{n_{2}}, \cdots,
\underbrace{\omega^{k-1}, \cdots \omega^{k-1}}_{n_{k}})$ with $\omega
= e^{2\pi i/k}$.  It is easy to see, however, that only the choice
$n_{1} = \cdots = n_{k} = F$ is left on the quantum moduli space.  For
any other choice at least one of the remaining $SU(n_{j})$ gauge
groups satisfies $n_{j} > F$, and hence a dynamical superpotential is
generated and the moduli space is lifted quantum mechanically.  For
the only possible choice $n_{1} = \cdots n_{k} = F$, the gauge group
is broken to $SU(F)^{k} \times U(1)^{k-1}$, where each $SU(F)$ factor
has $F$ flavors.  The dynamical scale $\Lambda^{(j)}$ of the
low-energy $j$-th $SU(F)$ theory is given by
\begin{equation}
\label{subscales}
    \Lambda^{(j)2F}
    = \Lambda^{2kF-F} 
    \frac{(h k (v\omega^{j-1})^{k-1})^{F}}{(k(v\omega^{j-1})^{(k-1)})^{2F}}
    = {\Lambda^{2kF-F} h^{F} 
\over k^F (v\omega^{j-1})^{(k-1)F}}
\end{equation}
where the combination $hv^{k-1}$ is the mass of the adjoint chiral
multiplet due to the superpotential coupling.  Here and hereafter, we
do not keep prefactors that depend on $O(1)$ numbers and $k$, but
retain only power dependencies.  Note that the dynamical scales of
$SU(F)$ factors are different in phase.  We have the quantum modified
moduli spaces for each $SU(F)$ factor:
\begin{equation}
    W = \sum_{j=1}^{k} X_{j} 
    ({\rm det} M^{(j)} - B^{(j)}\ov{B}^{(j)}-\Lambda^{(j)2F})
    = \sum_{j=1}^{k} X_{j} 
    \left({\rm det} M^{(j)} - B^{(j)}\ov{B}^{(j)}
    - \frac{h^{F}\Lambda^{2kF-F}}{k^F(v\omega^{j-1})^{(k-1)F}}\right),
    \label{eq:kmoduli}
\end{equation}
where $X_{j}$ are Lagrange multiplier fields.  The baryons $B^{(j)}$
and $\ov{B}^{(j)}$ are charged under the unbroken $U(1)^{k-1}$ gauge
groups.  The photons correspond to the gauge-invariants ${\rm Tr}
X^{j} W_{\alpha}$ for $j = 1, \cdots, k-1$.  There are $k F^{2}$ meson
degrees of  freedom and $2k$ baryons, subject to $k$ constraints.
Naively, this leaves $k F^{2} + k$ degrees of freedom, while,
according to the previous counting, at each point in the
$k-1$-dimensional space $\tr X^l$, $l=2,...,k$ there should be
$kF^2+1$.  However, this is not a contradiction because of
$U(1)^{k-1}$ gauge factors.  On a generic point on the moduli space,
the baryon fields break the $U(1)^{k-1}$ gauge group completely.
Therefore, $k-1$ of them are ``eaten'' by the Higgs mechanism.  The
number of chiral superfields in the low-energy limit is therefore
still $kF^{2}+1$.

Now the challenge is to study the limit $v \rightarrow 0$.  Clearly,
the moduli space described by Eq.~(\ref{eq:kmoduli}) is singular as $v
\rightarrow 0$.  We approach this limit in two different ways.

The first method is to approach $v \rightarrow 0$ when the
$U(1)^{k-1}$ factors are always broken.  It is then possible to
explicitly integrate out unnecessary degrees of freedom from
Eq.~(\ref{eq:kmoduli}).  The point is that $k-1$ degrees of freedom
out of baryons are ``eaten'' and hence the remaining degrees of
freedom can be parameterized by
\begin{equation}
    N^{(i)} \equiv B^{(i)} \ov{B}^{(i)},\qquad
    {\cal B} \equiv \prod_{i} B^{(i)},\qquad
    \ov{\cal B} \equiv \prod_{i} \ov{B}^{(i)},
\end{equation}
subject to a constraint
\begin{equation}
    {\cal B}\ov{\cal B} = \prod_{i} N^{(i)}.
\end{equation}
And the l.h.s. of the constraint is related to the original baryon 
operators by
\begin{equation}
    B\ov{B} = k^{kF} v^{k(k-1)F} (-\omega)^{{k(k-1)F\over 2}}{\cal
      B}\ov{\cal B} . 
\end{equation}
The latter relation is obtained in the following way.
The original baryon operator is defined as the determinant of the matrix
\be
\label{barmat}
\left(\begin{array}{cccc}
       Q^{(1)} & Q^{(2)}  &...  & Q^{(k)} \\
       (XQ)^{(1)} &  (XQ)^{(2)}  &...  & (XQ)^{(k)} \\
        .   &     &   &           \\
        .   &     &   &            \\
        .   &     &   &             \\
       (X^{k-1}Q)^{(1)} & (X^{k-1}Q)^{(2)}  &...  & (X^{k-1}Q)^{(k)} \\   
\end{array} \right)
\ee 
where each of the entries is an $F\times F$ matrix and the upper index
refers to the group of colors associated with the eigenvalue
$v\omega^i$ of $X$.  If we define the matrix $\Omega$ such that
$\Omega_{ij}={\cal I}_{F\times F} \ \omega^{(i-1)(j-1)}$ where ${\cal
  I}$ is the identity matrix, the determinant of the matrix in
eq.(\ref{barmat}) is $v^{k(k-1)F/2}\ ({\rm det} \Omega)\ {\cal B} $
with det $\Omega=k^{kF/2}\ (-\omega)^{{k(k-1)F\over 4}}$.

The $N^{(i)}$ are given directly from Eq.~(\ref{eq:kmoduli}) and we 
find
\begin{eqnarray}
    B\ov{B} &=& k^{kF} v^{k(k-1)F} (-\omega)^{{k(k-1)F\over 2}}
    \prod_{j} \left( {\rm det} M^{(j)}  
    - \frac{h^{F}\Lambda^{2kF-F}}{k^F(v\omega^{j-1})^{(k-1)F}} \right)
    \nonumber \\
    &=& \prod_{j} \left( k^F v^{(k-1)F}  (-\omega)^{{(k-1)F\over 2}}
      {\rm det} M^{(j)}  
    - (-)^{{(k-1)F\over 2}} h^{F}\Lambda^{2kF-F} \right).
    \label{eq:BovB}
\end{eqnarray}
Now by rewriting $M^{(j)}$ as
\begin{equation}
\label{submes}
    M^{(j)} = \frac{1}{k} \sum_{l=1}^{k} \frac{M_{l}}{v^{l-1}} 
    \omega^{-(j-1)(l-1)},
\end{equation}
we can smoothly take the limit
\begin{equation}
    v^{(k-1)} M^{(j)} \rightarrow M_{k} \omega^{-(j-1)(k-1)}
    = {M_{k} \over k}\omega^{j-1}.
\end{equation}
Then the constraint Eq.~(\ref{eq:BovB}) becomes
\begin{equation}
    B\ov{B} =  \prod_{j} (-1)^{{(k-1)F\over 2}}\left( {\rm det} M_{k}
    - h^{F}\Lambda^{2kF-F} \right).
\end{equation}
This is the quantum modified constraint among composites at $v
\rightarrow 0$, {\it as long as}\/ you approach the origin with all
the gauge groups always completely broken.  Note that the Higgs phase
and confining phase are equivalent in this theory because of quarks in
the fundamental representation.

What about taking the limit $v \rightarrow 0$ with keeping some or 
all of $U(1)$'s unbroken?  There are many reasons to believe that 
this limit leads to an interacting superconformal theory.  One way to 
see it is as follows.  We can always force the baryons to vanish in 
Eq.~(\ref{eq:kmoduli}), by adding a mass term to the quarks.  By 
adding a common mass term for simplicity,
\begin{equation}
    W = \sum_{j=1}^{k} X_{j} 
    \left({\rm det} M^{(j)} - B^{(j)}\ov{B}^{(j)}
    - \frac{h^{F}\Lambda^{2kF-F}}{k^F (v\omega^{j-1})^{(k-1)F}}\right)
    + m {\rm Tr} M_{1},
\end{equation}
and noting
\begin{equation}
    M_{1} = \sum_{j=1}^{k}  M^{(j)},
\end{equation}
we can solve $\partial W/\partial M^{(j)} = 0$ to find that $X_{j} 
\neq 0$.  This is enough to force all baryons to vanish.  Then we can 
ask the question what happens in the $v\rightarrow 0$ limit.  Because 
quarks are massive, we can integrate them out first instead, and add 
the superpotential $h {\rm Tr}X^{k+1}$ afterwards.  Once the quarks 
are integrated out, the theory is nothing but the $N=2$ Yang--Mills 
theory, whose curve is known.  Adding the superpotential is known to 
make the theory flow to an Argyres--Douglas fixed point.  It was 
worked out explicitly for the $SU(3)$ and $k=2$ case 
\cite{ad}, but it is believed that any $SU(N)$ theory with 
any $k$ would lead to such non-trivial fixed-point theories, as long 
as $k < N$.  Therefore for $F \geq 2$, the theory will flow to 
superconformal theories.  When $F=1$, however, the superpotential is 
(presumably) irrelevant, and the theory is given by the Coulomb branch 
of the entire $SU(N)$ $N=2$ Yang--Mills.  

\section{The dual magnetic theory with {\bf $kF=N$} }

For $F=kN$, the general strategy to study the magnetic theory is to
start from the $SU(k)$ theory with $F+1$ flavors, where the spectrum
is given by the fields in Table~2,
and then add a deformation 
\be
\label{massdef}
W_{def}=m (M_1)_{F+1,F+1}
\ee
corresponding to the mass term $m Q_{F+1}\QQbar_{F+1}$ of the electric theory.

We will begin by discussing a generic  $k$ case, and then we will consider
the special $k=2$ case, that presents some special subtleties.

\subsection{{\bf  $k\neq 2$}}

In the generic case $k\neq 2$
 the superpotential is  
\be
\label{wmagn}
W_{magn}=-{h \over k+1} {\rm Tr}Y^{k+1} + {h \over \mu^2} (M_1 \qbar Y^{k-1} q +...+M_k \qbar q).
\ee 
 
Pretty much like in the electric theory,
on the moduli
space $Y^k$ must be proportional to the identity.
In the following, we will analyze first the flat direction 
$Y^k\neq 0$ and then consider the special point   $Y^k= 0$.  

\subsubsection{{\bf $v^k={1\over k}$Tr$Y^k (={1\over N}$ Tr$X^k) \neq0$}}

In the points of the moduli space with $v^k={1 \over 2} Tr Y^k\neq 0$
the adjoint of the magnetic theory (like that of the electric theory)
is diagonalizable with a complex gauge transformation. 
In the electric theory with $F+1$ flavors, the only vacuum which
is stable under a deformation such as a mass term for the $F+1$-th flavor 
breaks $SU(N)$ to $SU(N/k)\times U(1)^{k-1}$,
and correspondingly in the magnetic theory $SU(k)$ is broken to
$U(1)^{k-1}$.  The baryons in the $k$ decoupled sectors of the
magnetic theory are
\be
\label{bark}
b^{(j)}_i={\Lambda^{(j)\ (F-1/2)}h^{{F+1\over k}}\over (\mu^{(j)})^{1/2}} 
q^{(j)}_i,\ \ i=1,F+1,\ \ j=1,k
\ee
where the $\Lambda^{(j)}$'s are the dynamical scales of the $k$
(electric) subsectors
see eq.(\ref{subscales})and the $\mu^{(j)}$'s
are related to the original scale $\mu$ by the relation 
\be
\label{submu}
\mu^{(i)}={\mu^2\over k (v\omega^{j-1})^{k-1} h}.
\ee 
The latter is obtained by matching the relation
$\Lambda_{el}^{2kF-F}\Lambda_{magn}^{-F}=\left({\mu \over
    h}\right)^{2F}$ of the high energy theory to the relation
$(\Lambda_{el}^{(i)})^{2F}(\Lambda_{magn}^{(i)})^{-F}
=(\mu^{(i)})^{F}$ of the low energy sectors.

The mesons $M^{(j)}$ of the subsectors are
\begin{equation}
\label{magsubmes}
    M^{(j)} = \frac{1}{k} \sum_{l=1}^{k} \frac{M_{l}}{v^{l-1}} 
    \omega^{-(j-1)(l-1)},
\end{equation}  

In each of the $k$  decoupled sector there are $F^2$ mesons, one baryon
and one anti-baryon subject to one constraint (which classically
doesn't involve the mesons). At a generic point of the moduli space,
the baryons break the residual $U(1)$ gauge factors, and the total
number of degrees of freedom is $kF^2+1$.

After decoupling the $Y$ degrees of freedom, the tree level 
superpotential is:
\be
W=\sum_{j=1,k} M^{(j)} {\qbar^{(j)} q^{(j)}\over\mu^{(j)} }. 
\ee 
However, as the non-Abelian factor of the magnetic gauge group is completely broken, instanton effects
have to be added to the above superpotential as matching
conditions.  After instanton effects are included, we expect the
superpotential to have the same form as that of the electric theory
with $F+1$ flavors ( i.e. $\left[{BMB\over \Lambda^{N-1}}-\det M\right]$
for each of the decoupled sectors), so that after massive fields
are integrated out the algebraic relations between baryons and mesons
are modified in such a way as to reproduce those of eq. (\ref{qmc1}).

The required instanton term is 
\be
\label{instk}
 \sum_j
-{{\rm det } M^{(j)}\over
\Lambda^{(j)\ (2F-1)}}
\ee
If we now add the deformation (\ref{massdef}), which,
in terms of the mesons of the decoupled sectors, reads
\be
m \sum_{j=1,k} M^{(j)},
\ee
the fields $M^{(1),...,()}_{F+1, i}$, $M^{(1),....,(k)}_{ i, F+1}$,
$i=1,F+1$, $q^{(1),...,(k)}_j$, $\qbar^{(1),...,(k)}_j$, $j=1,F$ become massive and can be 
integrated out. We can then  identify
\be
b^{(i)}\equiv {\Lambda^{(1)\ F-1/2} \over \mu^{(1)\ 1/2}} q^{(i)}_{F+1},
\ee
i.e. the operators $b^{(i)}_{F+1}$ in
eq.(\ref{bark}), 
and the analogous operators with $\qbar^{(i)}_{F+1}$,  with the baryons and
anti-baryons of the theories with zero colors,  and the constraints
between mesons and baryons turn out to be exactly like those of the
electric theory: 
\be {\rm det} M^{(i)}-
b^{(i)}\ov{b}^{(i)}=\tilde{\Lambda}^{(i)} 
\ee

\subsubsection{{\bf Tr$X^k$=Tr$Y^k=0$}}

In this case on the classical moduli space we have
\be
\qbar=(u' ,0,....,0),\ \  q=(0,0,...,0,u''), Y_{i,j}=\delta_{i+1,j}
u_i\ i,j=1,k 
\ee
with $u' u'' \Pi_{j=1,k} u_j=-m \mu^2$. Notice that, being the $u$'s
further constrained by $k-1$ nontrivial $D$-flatness conditions, only
one of the vevs is left unconstrained.  Adding the latter to the
$kF^2$ mesons we have $kF^2+1$ degrees of freedom, the same number as
in the electric theory for Tr$X^j=0$.

The instanton generated superpotential (\ref{instk}) reduces to
\bea
\label{insk}
&&W_{inst}=
-{
1\over k h^F \Lambda^{ 2kF-(F+1)}}\nn\\
 \sum_{l,m,..,z}
&& \delta_{l(k-l)+m(k-m)+...+z(k-z),k-1} 
\epsilon_{j_1,j_2,...j_{F+1}}(M_l)_{1,j_1}(M_m)_{2,j_2}...
(M_z)_{F+1,j_{F+1}}\nn\\ 
&&
\eea
In the (high-energy) theory with $F+1$ flavors, the $F$-flatness
conditions obtained by adding the new superpotential to that of
eq.(\ref{magn}) reproduce the (compositeness) constraints between
electric baryons and mesons.  Notice that the superpotential
(\ref{insk}) contains in particular the term
\be
-{M_1 {\rm cof} (M_k) \over h^F \Lambda^{2kF-(F+1)}}
\ee
which is crucial to obtain the quantum modified constraint in the low
energy theory.

Indeed, by integrating out the massive fields we find the quantum
generated constraint 
\be
\label{consk}
\la \det M_k\ra-            
{\Lambda^{2N-(F+1)}h^{F+1}\over \mu^2}
\la\qbar_{F+1}Y^{k-1} q_{F+1}\ra= h^F \Lambda^{2N-(F+1)} \times m=
h^F \tilde{\Lambda}^{(2N- F)}
\ee 
where $\tilde{\Lambda}$ is the scale of the low energy theory
with $F$ flavors. 

Being (for $v=0$) ${\cal M}$ of the
form

\be
\label{meson}
\left(\begin{array}{ccccc}
       M_1 & M_2 &...  & M_{k-1} & M_k\\
       M_2 & M_3 &...  & M_k     & 0  \\
        .   &     &   &    &        \\
        .   &     &   &    &        \\
        .   &     &   &    &         \\
       M_k  & 0   & 0 & 0 & 0    
\end{array} \right)
\ee
we find that it is 
\be
\label{mk}
(-\det{\cal M})^{1/k}=\det M_k.
\ee
Moreover, 
in the low energy theory with zero colors the product
of baryon and anti-baryon can be identified with 
\be
\label{bk}
b\ov{b}=(-)^{{k(k-1)F \over 2}}{\tilde{\Lambda}^{k(2N-F)}\over \mu^{2k}}h^{k(F+1)}
(\qbar_{F+1}Y^{k-1} q_{F+1})^k.
\ee
This can be understood in the following way.  Consider the magnetic
($SU(k)$) theory with $F+1$ flavors.  Among the
$k(F+1) \choose k$ baryons of the
magnetic theory there is 
\be
\label{ghostb}
  (-)^{{F+1\over 2}} {\Lambda^{  2N-(F+1) }h^{F+1} \over \mu^2} 
q_{F+1}(Yq_{F+1})..(Y^{k-1}q_{F+1})
\ee
(for the numerical factor see eq.(2.20)).
According to the duality vocabulary, this corresponds to the baryon 
\be
\label{barel} 
Q_1...Q_F(XQ_1)...(XQ_F)....(X^{k-1}Q_1)...(X^{k-1}Q_{F})
\ee
of the electric theory. After integrating out the $F+1$-th flavor, the
(opposite of the) latter is the only baryon remaining in the electric
theory, and, if the duality properties are preserved under mass
deformation, this should still correspond to the (opposite of the)
operator (\ref{ghostb}) in the magnetic theory.

Substituting eqs. (\ref{mk}) and (\ref{bk}) into eq.(\ref{consk}),  
we can rewrite the quantum modified constraint as
 
\be
\label{modcons}
(-)^{{k(k-1)F\over 2}}\la (\det{\cal M})\ra 
=(h^F \tilde{\Lambda}^{2N-F} (-)^{{(k-1)F\over 2}}\la (b\ov{b})^{1/k}\ra)^k  
\ee
which is in complete agreement with the quantum modified constraint
on the electric theory.

The constraint (\ref{modcons}) is also satisfied when $q=\qbar=Y=0$ and
only the mesons get a vev, in which case the SU(k) gauge symmetry is
unbroken and much of the above discussion seems not to hold.  Indeed,
it is not even obvious from the above considerations that this point
belongs to the moduli space.  The instanton superpotential is
generated if the gauge group is broken; if it was not, the
$F$-flatness condition would be the classical one, which is not
satisfied by $q=\qbar=Y=0$.  On the other hand this point can be
reached from the direction
$b^{(i)}=\ov{b^{(i)}}=0$ in the limit
$v\rightarrow 0$ and from the direction $v=0$,
${\rm det} M_k=h^F\Lambda^{2N-F}$, $B(\ov{B})=0$, $\ov{B}(B)\rightarrow 0$ on
the moduli space.  In this limit, $U(1)^{k-1}$ gauge invariance is unbroken
and additional charged massless fields can arise at singularities on
the moduli space where $SU(k)$ is recovered classically.  Therefore we
conclude that this limit is on the moduli space, where the theory
becomes superconformal.

\subsubsection{Explicit derivation of the instanton term}

The explicit derivation of the instanton term is very similar to the
case $k=2$. The 't Hooft effective vertex is given by 
\be
 \Lambda_{magn}^{2k-(F+1)}
 \tilde{q}^{F+1}\tilde{\qbar}^{F+1}\tilde{Y}^{2k}\lambda^{2k} 
\ee
In order to saturate the fundamental fermions, we use
thus the following couplings:

\begin{itemize}
\item ${h\over \mu^2}M_l \tilde{\qbar} Y^{k-l} \tilde{q}$ , ${h\over
    \mu^2}M_m \tilde{\qbar} Y^{k-m} \tilde{q}$,... for a total of
  $(F-1)$ times
\item${h\over \mu^2} \tilde{M_r} \qbar Y^{k-r} \tilde{q}$ once 
\item ${h\over \mu^2} \tilde{M_s} \tilde{\qbar} Y^{k-s}  q$ once
\end{itemize}
in such a way that the total power of $Y$ is $k-1$, to obtain
\be
\Lambda_{magn}^{2k-(F+1)}M_l...\tilde{M_r}\tilde{M_s}
\tilde{q}^2\tilde{\qbar}^2\tilde{Y}^{2k}\lambda^{2k} Y^{k-1} q \qbar 
{h^{F+1}\over \mu^{2(F+1)}}.
\ee
Then we use:
\begin{itemize}
\item $q^*\lambda\tilde{q}$ once
\item $\ov{q}^*\lambda\tilde{\ov{q}}$ once
\item $Y^*\lambda \tilde{Y}$ (2k-2) times
\item $-h Y^{k-1} \tilde{Y}^2$ once 
\end{itemize}
to end up with
\be
\Lambda_{magn}^{2k-(F+1)}M_l...\tilde{M_r}\tilde{M_s} 
{h^{F+2}\over \mu^{2(F+1)}}={1  \over h^F
\Lambda^{(2k-1)F-1 }}M_l...\tilde{M_r}\tilde{M_s} 
\ee
(where we used the relationship
$\Lambda^{(2K-1)F-1}\Lambda_{magn}^{(2K-1)-F}=\left({\mu\over
    h}\right)^{2F+2}$) which term can be embedded in the
superpotential term (\ref{insk}).  These are the only terms compatible
with all the symmetries, in particular the $U(1)_R$ symmetry.  
When
$\tr Y^k\neq 0$ this superpotential turns into the k terms in
eq.(\ref{instk}).
In this case there are no
symmetry reasons to forbid terms which mix different $\hat{M}_i$'s .  
The only reason is dynamical: as the $k$ sectors are
completely decoupled for $v\neq 0$, no term which mixes them can be
generated.
 
\subsection{{\bf$k=2$}}

As $k\neq 2$ case, we add one extra flavor to obtain the magnetic
$SU(2)$ theory, and decouple the extra flavor to find the quantum
modified moduli space.  However, the magnetic $SU(2)$ theory needs to
be augmented by an additional term in the superpotential to obtain the
same moduli spaces between the electric and magnetic theories as shown in
the Appendix.  

The superpotential of the $SU(2)$ theory is that of eq.(\ref{additional}),
to which we add a mass deformation for
the $F+1-$th flavor:    
\be
\label{wmagn}
W_{magn}={h \over \mu^2} (M_1 \qbar Y q +M_2 \qbar q) 
-\left({{\rm det} M^{(1)}\over  \Lambda^{(1)\ 3N/2-(F+1)}}+
{{\rm det} M^{(2)}\over  \Lambda^{(2)\ 3N/2-(F+1)}}\right)
+m (M_1)_{F+1,F+1}.
\ee
Apart from the
special choice of the superpotential, this is the theory considered in
\cite{isphases}.

With a real gauge transformation, we can put $q_{F+1}$ in the form
\be
\label{u1}
\ \  q_{F+1}  = \left(\begin{array}{c} 0 \\ u_2 \end{array}\right).
\ee
With this gauge choice, in order to satisfy both $F$-flatness and
$D$-flatness conditions, it must be

\be
\label{u23}
 \qbar_{F+1}  =(u_1,0),
\ \ \  Y =\left(\begin{array}{cc}
0 & u_3\\
u_4 & 0 \end{array}\right),
\ee
with $u_1 u_2 u_3= -m\mu^2$ (classically) and $D_3={u_1^{2}\over
  2}+{u_2^{2}\over 2} - (u_3^{2}+u_4^{2})=0$.  This leaves two out of
the four vevs $(u_1,u_2,u_3,u_4)$ unconstrained.  We also notice that
after $Y$ and $q_{F+1}$ get vev, the combinations $q_i (M_1)_{i,F+1}$,
$i=1,...,F$ and $q_i (M_2)_{i,F+1}$, $i=1,...,F+1$, become massive,
and can be integrated out.  The number of degrees of freedom is thus
$2F^2+2$, equal to that of the electric theory.
 
In particular $u_4$ can be either vanishing or not. In case it is not,
it is Tr$Y^2\neq 0$, $Y$ can be put (with a complex gauge
transformation) in diagonal form and the behavior of the theory has to
reproduce that of the electric theory for $v\neq 0$. We will analyze
this case first, and then discuss the case $u_4 = 0$ as the limit for
Tr$Y^2\rightarrow 0$.

\subsubsection{{\bf $v^2={1 \over 2}$ Tr$Y^2(={1\over N}$ Tr$X^2)\ \neq0$}}
When $u_4\neq 0$, $Y$ can be put in the diagonal form 
\be
\label{vy}
Y= \left(\begin{array}{cc}
      v & 0 \\
      0 & -v   \\ 
      \end{array}\right),
\ee   with $v=\sqrt{u_3 u_4}$,   
with a complex gauge transformation.

The theory is split in two decoupled subsectors and the superpotential can be written as:
\bea
\label{wmag1}
W_{mag1}&=&\left({1\over \mu^{(1)}}
\qbar^{(1)}  q^{(1)} M^{(1)}+{1\over \mu^{(2)}} 
\qbar^{(2)} q^{(2)}M^{(2)}\right)\nn\\
&-&\left({{\rm det} M^{(1)}\over  \Lambda^{(1)\ 3N/2-(F+1)}}+
{{\rm det} M^{(2)}\over  \Lambda^{(2)\ 3N/2-(F+1)}}\right)
+m (M^{(1)}_{F+1,F+1}+M^{(2)}_{F+1,F+1})
\eea
where  $\mu^{(1),(2)}=\pm {\mu^2\over 2vh}$, consistently with the general formula (\ref{submu}).

$\hat{M}_{1,2}$ can be identified with the mesons 
of the two decoupled sectors, while the baryons are
 
\be
{\Lambda^{(1)\ F-1/2} \over \mu^{(1)\ 1/2}} q^{(1)}_i,\ 
\ \ {\Lambda^{(2)\ F-1/2} \over \mu^{(2)\ 1/2}} q^{(2)}_i
\ee
where $\Lambda^{(1),(2)}$ are the dynamical scales of the two 
{\em electric} subsectors
and the anti-baryons the analogous operators with $\qbar$.

In each of the two decoupled sector there are $F^2$ mesons, one baryon
and one anti-baryon subject to one constraint (which classically
doesn't involve the mesons). At a generic point of the moduli space,
the baryons break the residual $U(1)$ gauge group, and the total
number of degrees of freedom is $2F+1$.

         The magnetic gauge group is broken to        
$U(1)$, and we might expect that 
instanton effects had to be added, but it turns out that
in the absence of the superpotential term involving ${\rm Tr}Y^{k+1}$
such effects vanish.

After massive fields ($M^{(1),(2)}_{F+1, i}$, $M^{(1),(2)}_{ i, F+1}$,
$i=1,F+1$, $q^{(1),(2)}_j$, $\qbar^{(1),(2)}_j$, $j=1,F$) are
integrated out, we can identify
\be
b^{(1)}\equiv {\Lambda^{(1)\ F-1/2} \over \mu^{(1)\ 1/2}} q^{(1)}_{F+1},
\ \ \ b^{(2)}\equiv {\Lambda^{(2)\ F-1/2} \over \mu^{(2)\ 1/2}} q^{(2)}_{F+1}
\ee
and the analogous operators with $\qbar_{F+1}$ with the baryons and
anti-baryons of the theories with zero colors and the constraints
between mesons and baryons turn out to be exactly like those of the
electric theory: 
\be {\rm det} M^{(1),(2)}-
b^{(1),(2)}\ov{b}^{(1),(2)}=\tilde{\Lambda}^{(1),(2)} 
\ee 
where $\tilde{\Lambda}^{(1),(2)}$ are the scales of the low energy
theories (with $F$ flavors) for the two decoupled sectors, related to the high
energy scales by $\tilde{\Lambda}^{(1),(2)\ 2F}= m \Lambda^{(1),(2)\ 
  (2F-1)}$.

\subsubsection{{\bf Tr$Y^2$(=Tr$X^2)=0$}}

In the limit $v=0$, the additional superpotential term in (\ref{additional}) reduces to
\be
\label{inst2}
W_{add}=-\left({ M_1 {\rm cof}(M_2)\over h^F \Lambda^{2N-(F+1)}}\right).
\ee
In the presence of this term, the low energy theory, containing only
the (now $F\times F$) meson matrices $M_1$ and $M_2$ and another
degree of freedom out of the parameters $(u_1,u_2,u_3)$ of
eqs.(\ref{u1}),(\ref{u23}), is characterized by the (quantum
generated) constraint

\be
\label{constr}
\la \det M_2\ra-            
{\Lambda^{2N-(F+1)} h^{F+1} \over \mu^2}
\la\qbar_{F+1}Y q_{F+1}\ra= \Lambda^{2N-(F+1)}m h^F=
h^F \tilde{\Lambda}^{(2N- F)}
\ee
where $\tilde{\Lambda}^{(2N- F)} = m \Lambda^{(2N-(F+1))}$ is the
scale of the low energy theory with $F$ flavors.

Indeed, in the low-energy theory, the operator corresponding to the
product of the baryon and the anti-baryon in the electric theory is
\be
\label{baryon}
 b\ov{b}\equiv (-)^F {\Lambda^{2(3F-1)} h^{2(F+1)} \over \mu^4} (
 \qbar_{F+1}Y q_{F+1} )^2=(-)^F 
{\tilde{\Lambda}^{6F} h^{2(F+1)}\over m^2\mu^4} ( \qbar_{F+1}Y q_{F+1} )^2.
\ee
which is the specialization to the case $k=2$ of eq.(\ref{bk}).

Therefore, the moduli space for $N=2F$ is characterized by the constraint
\be
\label{qmc}
\la (-)^{F} {\rm det} {\cal M}\ra 
=\left(h^F\tilde{\Lambda}^{(2N- F)}+ (-)^F \sqrt{\la b\ov{b}\ra}\right)^2\  .
\ee
and this agrees with the quantum modified constraint
on the electric moduli space.

At the singular point $M_2=h^F
\tilde{\Lambda}^{2N-F}$, $b=\ov{b}=0$, where the $SU(2)$ gauge group
is unbroken.  Dynamics there must be superconformal, which fact is also consistent 
with the suggestion of ref.~\cite{isphases}.

\section{Conclusion}

In this paper, we have investigated the dynamics of supersymmetric
gauge theories with less flavor than the Kutasov-duals.  By
integrating out a flavor from the known duality pair, we demonstrated
that the theory has a quantum-modified moduli space, as expected from
the analogy to the Seiberg duality in $SU(N)$ QCD.  However, a point
on the moduli space becomes superconformal, a distinct behavior from
the Seiberg duality.  In fact, such a behavior had been seen in an
isolated example of $SU(2)$ gauge theory with two doublets and a
triplet, and our result generalizes this behavior to a whole class of
theories.  
We also pointed out that in the case $k=2$, $N=2F-2$, the magnetic theory
is not capable to reproduce the structure of the moduli space in
the electric theory unless we supply the magnetic superpotential with
an additional term, that was neglected in previous literature.

This result gives additional information on the phase space of these
theories.

\appendix

\renewcommand{\thesection}{Appendix \Alph{section}}
\section{Non-trivial Checks of Kutasov duality\label{eq:appendix}}
\renewcommand{\theequation}{\Alph{section}.\arabic{equation}}
\setcounter{equation}{0}
\setcounter{footnote}{0}

In this section we present additional new non-trivial checks of Kutasov duality,
showing in particular that in the case $N=2F-2$ the additional superpotential term in 
(\ref{additional}) has indeed to be added in the magnetic theory. 
The question is if moduli spaces agree between the electric and
magnetic theories.  This check had not been done explicitly in
literature to the best of our knowledge.  We focus on the case $N =
k(F-n)$, $n>0$ because it is relevant to our discussion of the
quantum modified moduli space when $N=kF$ ($n=0$).  

In the electric theory, the dressed quarks form a $N\times kF$ matrix,
and hence the dressed meson matrix ${\cal M}$ has a rank less than or
equal to $N = k(F-n) < kF$.  In this case, the following classical
constraint holds
\begin{equation}
\label{classconstr}
  ({\cal M})_{i_1}^{j_1} ({\cal M})_{i_2}^{j_2} \cdots 
  ({\cal M})_{i_{F-n}}^{j_{F-n}}
  \epsilon^{i_1 i_2 \cdots i_{F-n} i_{F-n+1} \cdots i_F} 
  \epsilon^{j_1 j_2 \cdots j_{F-n} j_{F-n+1} \cdots j_F}
  = B^{i_{F-n+1} \cdots i_F} \bar{B}_{j_{F-n+1} \cdots j_F}.
\end{equation}
Here, the indices run over $kF$ values.  In particular, when not along
the flat direction ${\rm Tr}X^k \neq 0$, the only surviving piece in
the left-hand side is given by $M_k$.  Therefore,
\begin{eqnarray}
  \lefteqn{
  \left[(M_k)_{i_1}^{j_1} (M_k)_{i_2}^{j_2} \cdots 
  (M_k)_{i_{F-n}}^{j_{F-n}}
  \epsilon^{i_1 i_2 \cdots i_{F-n} i_{F-n+1} \cdots i_F} 
  \epsilon^{j_1 j_2 \cdots j_{F-n} j_{F-n+1} \cdots j_F}\right]^k
  } \nonumber \\
  &=& B^{i_{F-n+1} \cdots i_F; i_{F-n+1} \cdots i_F; \cdots i_{F-n+1}
    \cdots i_F} 
  \bar{B}_{j_{F-n+1} \cdots j_F; j_{F-n+1} \cdots j_F; \cdots
    j_{F-n+1} \cdots j_F}. 
\end{eqnarray}
Here, the indices run only over the original flavors 1--$F$.

The same constraint is reproduced in the magnetic $SU(kn)$ theory due
to $k$-instanton effect.  When ${\rm rank}M_k = F-n$, we can integrate
out $F-n$ dual quarks from the theory such that the low-energy
dynamical scale is given by
\begin{eqnarray}
  \Lambda_{LE\ magn}^{2kn-n} = \Lambda_{magn}^{2nk-F}
  \widehat{\rm det} \left(\frac{h}{\mu^2} M_k\right).
\end{eqnarray}
$\widehat{\rm det}$ is the non-vanishing minor determinant for the rank
$F-n$ matrix.  There is a unique baryon operator 
\begin{equation}
  b = q^k (Yq)^k \cdots (Y^{k-1} q)^k
\end{equation}
in the low-energy $SU(kn)$ theory with $n$ flavors, and it breaks the
gauge group completely.  Then the instanton effects need to be
considered.  The $k$-instanton background gives $k$ zero modes to each
flavor, $2nk^2$ zero modes to both gaugino and $\tilde{Y}$.  Along the
baryon (and anti-baryon) direction, each quark zero modes combine with one 
gaugino zero mode  to give
the corresponding  squark VEV; the remaining gaugino zero modes combine with
$2nk(k-1)$ of the  $\tilde{Y}$ ones to give $2nk(k-1)$ powers of $Y$; $nk(k-1)$ of the latter
combine with the remaining  
$\tilde{Y}$ fermions
giving a factor $h^{nk}$.  Therefore, the instanton background gives
the correlation function
\begin{eqnarray}
  (q_{F-n+1})^k \cdots (q_{F})^k
  (\bar{q}_{F-n+1})^k \cdots (\bar{q}_{F})^k
  Y^{nk(k-1)}
  = h^{nk} \Lambda_{LE\ magn}^{k(2kn-n)}.
\end{eqnarray}
The left-hand side is nothing but $b\bar{b}$ for the remaining $n$
flavors.  We rewrite this result in terms of the electric scale using
the matching condition
\begin{equation}
  \Lambda^{2N-F} \Lambda_{magn}^{2\tilde{N}-F}
  = \left( \frac{\mu}{h} \right)^{2F}.
\end{equation}
We find
\begin{eqnarray}
  b\bar{b}
  &=& h^{nk} \left( \frac{h}{\mu^2} \right)^{k(F-n)}
  (\widehat{\rm det}M_k)^{k} \frac{1}{\Lambda^{k(2N-F)}} 
  \left( \frac{\mu}{h} \right)^{2kF} \nonumber \\
  &=& \mu^{2kn} h^{-kF} \frac{1}{\Lambda^{k(2N-F)}} (\widehat{\rm det}M_k)^{k}.
\end{eqnarray}
Matching between baryon operators is
\begin{equation}
  B = h^{kF/2} \mu^{-\tilde{N}} \Lambda^{k(2N-F)/2} b
\end{equation}
for appropriate flavor indices.  Therefore, the factors in $h$, $\mu$,
and $\Lambda$ all work out to give
\begin{equation}
  B\bar{B} = (\widehat{\rm det}M_k)^{k}
\end{equation}
for the relevant flavor combination.  

Along the flat direction $v\neq 0$, the degrees of freedom of the adjoint
field are integrated out, and in each of the $k$ decoupled 
low-energy sectors we have some classical constraints between 
mesons and baryons analogous to those of eq.(\ref{classconstr}).
If for every subsector it is $N^{(i)} < F+1$, every subsector of the corresponding
magnetic theory has a residual non-Abelian gauge group, whose
dynamics is known from  Refs. \cite{dual} and has been found to reproduce the
constraints of the electric theory.
If for some subsectors it is either $N^{(i)}=F$ or $N^{(i)}=F+1$, in the corresponding
magnetic subsector the (non-Abelian part of the) gauge group is completely broken,
and its superpotential must include  instanton contributions from the
theory with the adjoint.
Take as an example the case $k=4$, $n=1$, and consider the flat direction where $X$ 
has $F-2$ eigenvalues $v$ and $-v$ and $F$ eigenvalues $iv$ and $-iv$.
In the magnetic theory we have two sectors (which we will label (3) and (4) )
in which the gauge group is purely Abelian.
The one-instanton background gives 8 zero modes for gauginos and $\tilde{Y}$
and $F$ zero modes for the squarks. Following the very same steps as in subsection 4.1.3,
it is easy to see that the terms
\be
-\left({{\rm det} M^{(3)}\over  \Lambda^{(3)\ 2F}}+
{{\rm det} M^{(4)}\over  \Lambda^{(4)\ 2F}}\right)
\ee
are generated.

But when the magnetic group is
$SU(2)$, $k=2$, $n=1$, there is no superpotential in
the magnetic theory ${\rm Tr}Y^{k+1} = {\rm Tr}Y^3 = 0$.  Because this
superpotential term was crucial for the instanton effect to reproduce
the classical constraint in the electric theory, the dual theory
without the ${\rm Tr}Y^3$ term does not describe the same moduli
space.  The only way to remedy it is to introduce an additional term
to the superpotential
\begin{equation}
  W_{inst}=-\left({{\rm det} M^{(1)}\over  \Lambda^{(1)\ 2F}}+
{{\rm det} M^{(2)}\over  \Lambda^{(2)\ 2F}}\right)
\end{equation}
that reduces to
\begin{equation}
  W_{inst}=-\left({ M_1 {\rm cof}(M_2)\over h^F \Lambda^{2N-F}}\right).
\end{equation}
for $v=0$.

As shown in section 4.2,             
this is sufficient 
to reproduce the constraints of the electric theory.
Note that this additional term in the superpotential is a
one-instanton contribution in the magnetic theory because we can
rewrite it as
\begin{equation}
  W_{inst}=- \frac{h}{\mu^2} M_1 {\rm cof}\left(\frac{h}{\mu^2}
    M_2\right)
  \Lambda_{magn}^{4-F}.
\end{equation}


\begin{thebibliography}{99}

\bibitem{dual}
N.~Seiberg,
``Electric - magnetic duality in supersymmetric nonAbelian gauge theories,''
Nucl.\ Phys.\ B {\bf 435}, 129 (1995)
[hep-th/9411149].
\bibitem{kut1}
D.~Kutasov,
 ``A Comment on duality in N=1 supersymmetric nonAbelian gauge theories,''
Phys.\ Lett.\ B {\bf 351} (1995) 230
[hep-th/9503086].
\bibitem{kut2}
D.~Kutasov and A.~Schwimmer,
 ``On duality in supersymmetric Yang-Mills theory,''
Phys.\ Lett.\ B {\bf 354} (1995) 315
[hep-th/9505004].
\bibitem{kut3}
 D.~Kutasov, A.~Schwimmer and N.~Seiberg,
 ``Chiral Rings, Singularity Theory and Electric-Magnetic Duality,''
Nucl.\ Phys.\ B {\bf 459} (1996) 455
[hep-th/9510222]. 
\bibitem{more}
K.~A.~Intriligator,
``New RG fixed points and duality in supersymmetric SP($N_c$) and
SO($N_c$) gauge theories,'' 
Nucl.\ Phys.\ B {\bf 448}, 187 (1995)
[arXiv:hep-th/9505051];\\
R.~G.~Leigh and M.~J.~Strassler,
``Duality of Sp(2$N_c$) and S0($N_c$) supersymmetric gauge theories with
adjoint matter,'' 
Phys.\ Lett.\ B {\bf 356}, 492 (1995)
[arXiv:hep-th/9505088].
K.~A.~Intriligator, R.~G.~Leigh and M.~J.~Strassler,
``New examples of duality in chiral and nonchiral supersymmetric gauge
theories, 
Nucl.\ Phys.\ B {\bf 456}, 567 (1995)
[arXiv:hep-th/9506148].
\bibitem{discrete}
C.~Csaki and H.~Murayama,
``Discrete anomaly matching,''
Nucl.\ Phys.\ B {\bf 515}, 114 (1998)
[arXiv:hep-th/9710105].
\bibitem{cm}
C.~Csaki and H.~Murayama,
``New confining N = 1 supersymmetric gauge theories,''
Phys.\ Rev.\ D {\bf 59} (1999) 065001
[hep-th/9810014].
\bibitem{ad}
P.C. Argyres and M.R. Douglas,
``New Phenomena in $SU(3)$ Supersymmetric Gauge Theory,''
Nucl.\ Phys.\ {\bf B448} ((1995) 93
[hep-th/9505062].
\bibitem{is}
K.~Intriligator and N.~Seiberg,
 ``Lectures on supersymmetric gauge theories and electric-magnetic  duality,''
Nucl.\ Phys.\ Proc.\ Suppl.\ {\bf 45BC} (1996) 1
[hep-th/9509066].
\bibitem{isphases}
K.~Intriligator and N.~Seiberg,
``Phases of N=1 supersymmetric gauge theories in four-dimensions,''
Nucl.\ Phys.\ B {\bf 431}, 551 (1994)
[hep-th/9408155].
\bibitem{instanton}
C.~Csaki and H.~Murayama,
``Instantons in partially broken gauge groups,''
Nucl.\ Phys.\ B {\bf 532}, 498 (1998)
[arXiv:hep-th/9804061].

\end{thebibliography}
\end{document}